# Cryogenic Neuromorphic Hardware

Md Mazharul Islam[1], Shamiul Alam[1], Md Shafayat Hossain[2], Kaushik Roy[3] and Ahmedullah Aziz[1*]

[1]Dept. of Electrical Eng. & Computer Sci., University of Tennessee, Knoxville, TN, 37996, USA
[2]Dept. of Physics, Princeton University, Princeton, NJ, 08544, USA
[3]Dept. of Electrical and Computer Engineering, Purdue University, West Lafayette, IN, 47906, USA
*Corresponding Author's Email: aziz@utk.edu

**Abstract-** The revolution in artificial intelligence (AI) brings up an enormous storage and data processing requirement. Large power consumption and hardware overhead have become the main challenges for building next-generation AI hardware. To mitigate this, Neuromorphic computing has drawn immense attention due to its excellent capability for data processing with very low power consumption. While relentless research has been underway for years to minimize the power consumption in neuromorphic hardware, we are still a long way off from reaching the energy efficiency of the human brain. Furthermore, design complexity and process variation hinder the large-scale implementation of current neuromorphic platforms. Recently, the concept of implementing neuromorphic computing systems in cryogenic temperature has garnered intense interest thanks to their excellent speed and power metric. Several cryogenic devices can be engineered to work as neuromorphic primitives with ultra-low demand for power. Here we comprehensively review the cryogenic neuromorphic hardware. We classify the existing cryogenic neuromorphic hardware into several hierarchical categories and sketch a comparative analysis based on key performance metrics. Our analysis concisely describes the operation of the associated circuit topology and outlines the advantages and challenges encountered by the state-of-the-art technology platforms. Finally, we provide insights to circumvent these challenges for the future progression of research.

## I. Introduction

With the advent of artificial intelligence as one of the leading decision-making pathways in modern technology, it is getting harder to process and store an enormous amount of data in conventional von Neumann architecture [1], [2]. Moreover, the evolution of microelectronic circuits for almost half a century has reached a fundamental bottleneck where the miniaturization of electronic devices has become challenging due to several practical concerns [3]. The massive expansion of data requires excessive storing and processing energy. For example, the US data centers consume power in the range of terawatt-hour (TWh) [4]. Additionally, several device reliability issues emerge as a major concern for short-channel devices. These unassailable challenges have led us to look for a fundamentally new computational architecture. From neuroscience, we know that the human brain can process an enormous amount of data only by consuming around 20 W of power [5]. Also, our brain can perform complex tasks operating at an ultra-low frequency (few Hertz), thanks to the high degree of parallelism in its architecture [6]. So, it is intuitive to try to mimic the biological brain structure to build a novel architecture and on this basis, neuromorphic computing has emerged [7]. In our brain, information is transmitted in the form of spikes between neurons and synapses [8], [9]. Imitation of this biomimetic behavior has given rise to the spiking neural network (SNN) where the information is processed through neurosynaptic spiking dynamics [10]–[12].

Although the existing software implementations of spiking neural networks are precise and flexible, they are expensive in terms of computation and power efficiency [13]–[15]. Keeping this in mind, researchers have been trying to build a dedicated hardwarse platform for processing SNN-based algorithms. Several device platforms based on memristor, magnetic materials, phase change materials (PCM), and complementary metal-oxide-semiconductor (CMOS) have been explored for this purpose. SNN implementation based on these devices relies on the intrinsic dynamics of the devices to realistically emulate biological behavior [16]–[26]. However, they are still far behind the human brain in terms of power efficiency and compactness. Also, process variation has become a major concern for large-scale architecture implementation [27]. For these challenges, researchers are constantly exploring novel platforms with higher energy efficiency and scalability.

The primary source of power consumption in a large neuromorphic network is the massive interconnectivity between the neurons and synapses [28]. In this regard, superconductors and superconducting devices have lossless characteristics which can be employed as low-power interconnects in a neuromorphic network [29]. Besides, superconducting devices exhibit switching characteristics with unprecedented power consumption and ultra-high speed. There are also several non-superconducting materials such as $NbO_x$ which exhibit neuromorphic characteristics in cryogenic temperature [30]. The extraordinary power consumption of these devices indicates that cryogenic devices can open a new era of next-generation ultra-low power and compact neuromorphic hardware platform.

In this review, we discuss state-of-the-art cryogenic neuromorphic hardware. We organize our discussions as the following: in section II, we discuss the motivation for exploring cryogenic neuromorphic hardware and the primary advantages of these architectures. In section III, we shed the light on different cryogenic neuromorphic hardware where both superconducting and non-superconducting designs are analyzed. In section IV, we present a comparative analysis among different methodologies and discuss the major challenges faced by different cryogenic neuromorphic platforms. Finally, we conclude our discussion with several research prospects.

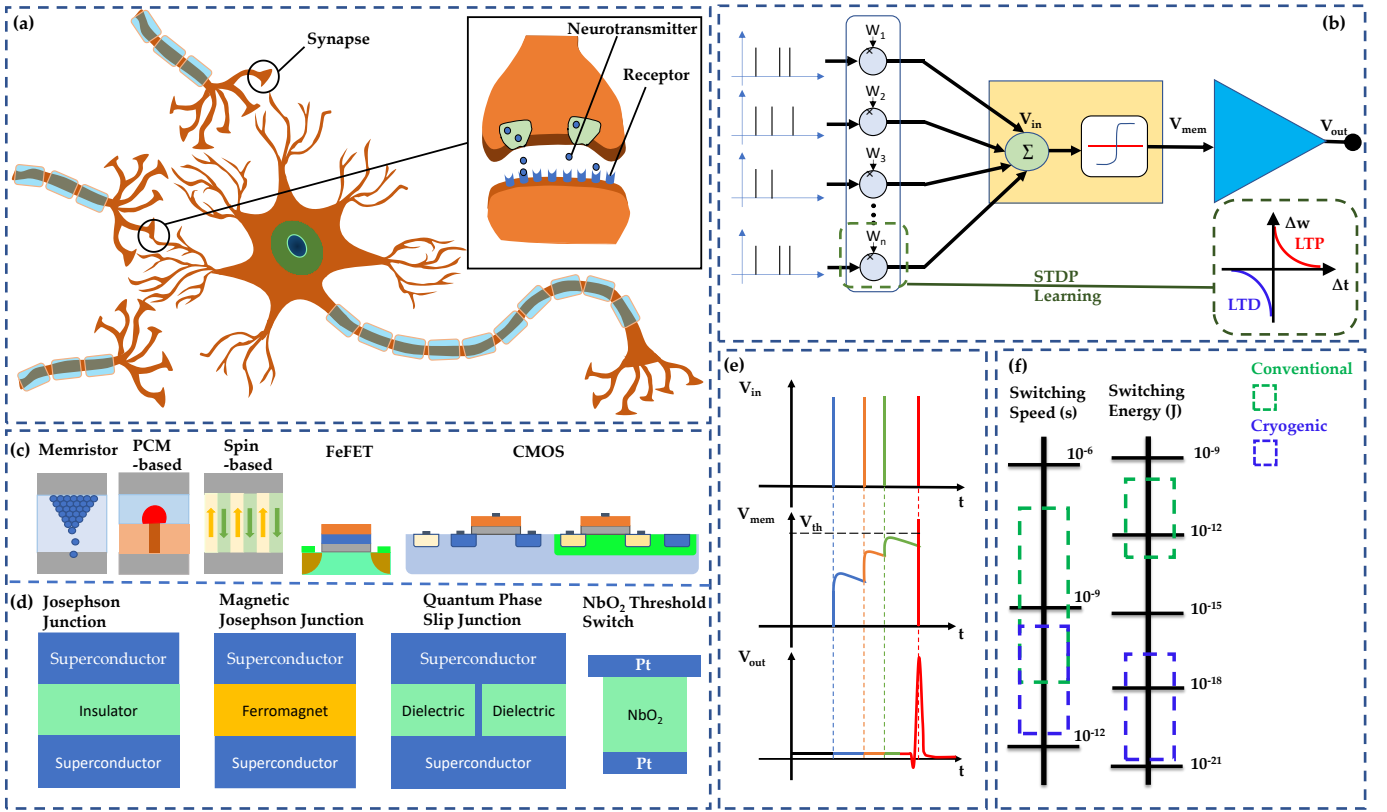

**Figure 1:** **(a)** Biological neuron connected with multiple neurons through synapses, inset shows the transportation of neurotransmitter. **(b)** Electronic model of a neuromorphic system showing the integration of weighted spikes. **(c)** Several conventional hardware platforms. **(d)** Several cryogenic platforms for neuromorphic hardware. **(e)** Input spikes ($V_{in}$), corresponding membrane potential ($V_{mem}$), and output spike ($V_{out}$) of a leaky integrating and fire (LIF) neuron. An output spike is generated after $V_{mem}$ crosses the threshold voltage ($V_{th}$). **(f)** Switching speed and switching energy comparison of conventional and cryogenic hardware [28].

## II. Why Cryogenic Neuromorphic?

For an efficient neuromorphic network, the network system should have sufficient power efficiency, bio-realism, and compactness compared to the human brain. The human brain has several distinct features that make it different from a computer. In the core of our brain's building block, neurons, synapses, axons, and dendrites are the most significant components (see fig. 1(a) and fig. 1(b) for the biological and electronic representation of a neuromorphic system, respectively). Neurons are spiking element which generates spikes and synapses control the signal transport between neurons through adjustable connectivity strength. Fig. 1(e) illustrates the behavior of a spiking neuron. There are 86 billion neurons in the human brain and each one is connected to thousands of others through synapses [31]. This massive interconnectivity enables our brain to be compact (around 1.5 kg in weight and 1260 $cm^3$ in volume) and highly energy-efficient (20 W of power consumption) [32]–[34]. Thus, the neuromorphic platform should be less dissipative so that a high degree of parallelism can be achieved. So far, researchers have sought a wide variety of devices as neuromorphic hardware platforms in the last few decades. There are several challenges associated with the conventional platforms that limit their performance in practical implementation.

(i) Dissipative interconnection limits the large-scale integration of conventional hardware [35]. The resistive interconnect loss results in signal latency and inefficiency.
(ii) Bio-realism is a crucial requirement for a network to be brain-like [36]–[38]. Inherent characteristics of a device determine the degree of bio-realism that can be achieved through a careful design approach[39]. Although conventional neuromorphic hardware exhibits a varying degree of bio-realism, most of them lack design simplicity. A simpler design implies fewer hardware resources and a lower area for each component.
(iii) Another bottleneck of conventional hardware platforms is their operating frequency. For high-frequency applications, the switching time should be sufficiently fast [40]. Most of the conventional platforms are limited in terms of the highest operable frequency [28].
(iv) Each of the technology platforms has its own individual drawbacks. Although CMOS devices can operate at very low temperatures (4K), several reliability issues emerge at low temperatures [41]–[43]. For example, due to the hot-carrier effects, the device lifetime decreases significantly [44]. Moreover, it is challenging to design CMOS circuits with biological realism due to massive design complexity. Resistive switching-based devices such as memristors and MTJ-based neuromorphic devices are being widely explored for their high scalability [10],[45]. But at low temperatures, resistance variation occurs due to the geometric fluctuations of the filament in the memristor [46]. Also, memristive devices suffer from low endurance, and high latency, and are prone to process variation whereas MTJ suffers from a low ON/OFF ratio and low density [16], [47], [48].

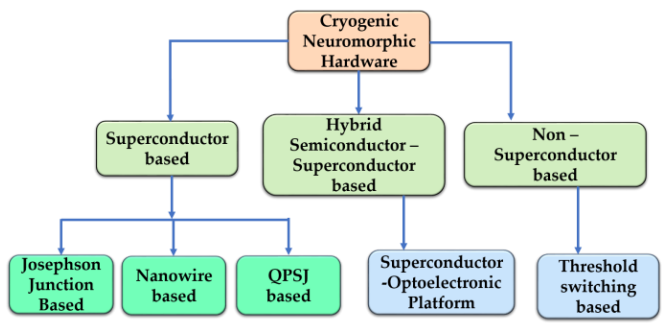

**Figure 2:** Hierarchical classification of the state-of-the-art cryogenic neuromorphic hardware.

Because of these challenges, it is challenging to build an efficient neuromorphic network on conventional platforms. Therefore, a novel platform that is capable of tackling these challenges can pave the way for the next-generation neuromorphic hardware platform.

Superconducting materials and devices are well known for their lossless characteristics and therefore they are an obvious choice of exploration for designing efficient neuromorphic circuits. Josephson junction (JJ), quantum phase slip junction (QPSJ), and superconducting nanowire (SNW) are some of the devices that have been used extensively in single-flux quantum (SFQ) and rapid single-flux quantum (RFSQ) circuits (fig. 1(d)). These devices consume orders of magnitude lower switching energy compared to the conventional non-cryogenic devices [49]–[53]. Besides, ultrafast switching behavior in these devices enables them to operate in hundreds of GHz of frequency [50], [54]. Apart from the superconducting devices, non-superconducting devices such as $NbO_x$ also exhibit fast switching characteristics in low temperatures for which they have been reported in the cryogenic neuromorphic circuit generating high-frequency oscillation (several GHz) [55]. Also, the inherent characteristics of these devices are suitable for designing bio-realistic neuromorphic circuits with low design complexity and fewer hardware resources. Fig. 1(f) shows the comparison of speed and energy between the conventional and cryogenic neuromorphic platforms.

Moreover, the operating temperature of these devices offers the inherent advantage of optimum performance in low-temperature applications. For example, the peripheral control of quantum computers can highly benefit from a neuromorphic circuit in cryogenic temperature [56]. The parameters of the quantum circuit can be trained by the neuromorphic network. Marković *et al.* have demonstrated a neural network that has been implemented in brain-inspired quantum hardware to accelerate the computation process [56]. Also, a cryogenic environment offers much higher noise immunity compared to the room temperature [57]. This is a significant advantage since noise can affect the overall performance of the electrical oscillator used in a neuromorphic system [18], [58]. Lastly, cryogenic devices have excellent thermal properties required for aerospace applications [59]. These unique advantages of cryogenic devices have incited the exploration of neuromorphic hardware in the cryogenic domain.

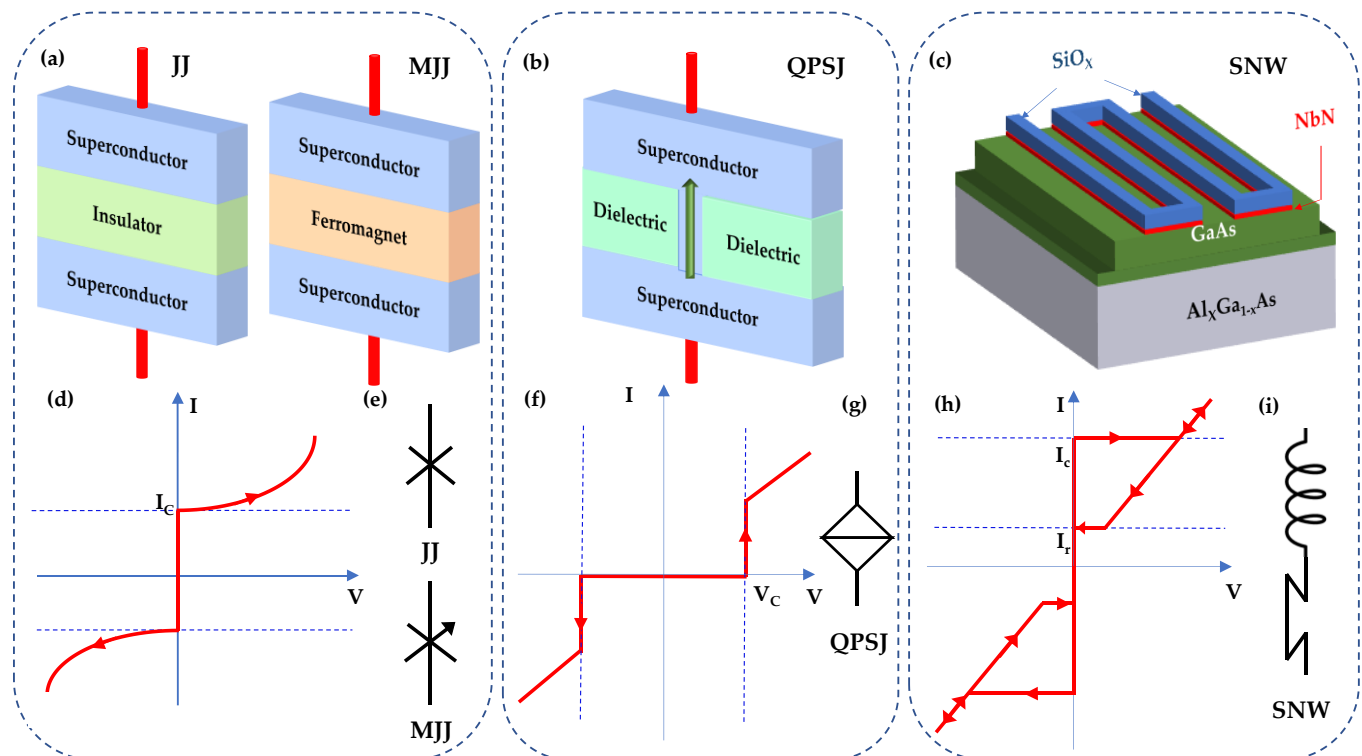

**Figure 3:** Structure of **(a)** an SIS Josephson junction (JJ) and magnetic Josephson junction (MJJ), **(b)** a quantum phase slip junction (QPSJ), and **(c)** a superconducting nanowire (SNW). **(d)** I-V characteristics and **(e)** electrical symbol of the JJ and the MJJ. **(f)** I-V characteristics and **(g)** electrical symbol of the QPSJ. **(h)** I-V characteristics and **(i)** electrical symbol of the SNW.

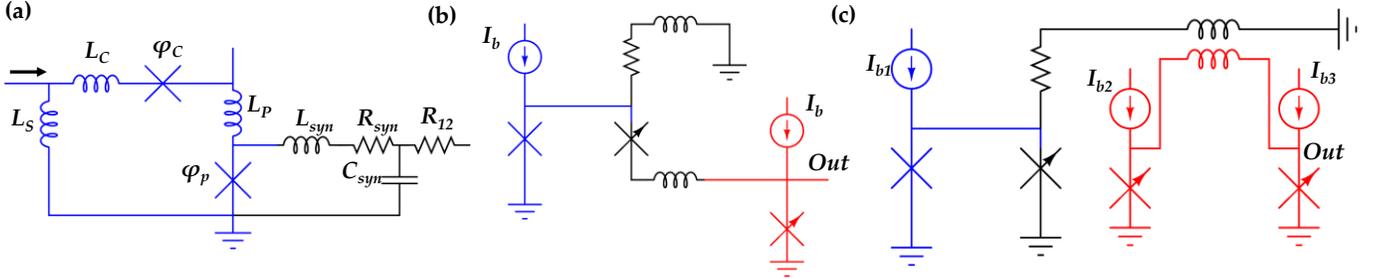

**Figure 4:** Three different Josephson junction-based neuromorphic circuit topologies proposed in [39], [64] and [65]. *Blue, black,* and *red* colors signifie pre-synaptic neuron, synapse, and post-synaptic neuron.

### III. State-of-the-art Cryogenic Neuromorphic devices

So far, several cryogenic hardware has been proposed for designing neurons and synapses in a neuromorphic network. We primarily classify them based on the core cryogenic device. The classification scheme is outlined in Fig. 2.

*III.A. Superconductor-based Neuromorphic Hardware*

We start with superconducting device-based neuromorphic designs. Several superconducting devices such as JJ, QPSJ, and SNW have been utilized to design energy-efficient neuromorphic circuits. Figure 3 demonstrates the physical structure of these devices and their typical *I-V* characteristics. These devices manifest abrupt switching that is suitable for the spike generation process. As important as the switching characteristics, a neuromorphic architecture must have the ability to transport spikes over a long distance without any significant power dissipation. In a superconducting transmission line, the resistive loss is not present, and the dispersion is low up to a certain frequency. This frequency is known as the gap frequency, $f_g (= 2\Delta_{sc}/h)$ [60]. In most of the superconducting circuits, Nb transmission line is used for which the gap frequency is reported as much as 650 GHz [60]. It is reported that a single flux quantum (SFQ) pulse with the duration of 1 ps can travel up to 10 mm distance without significant loss and dispersion [60][61]. If the inherent power consumption is considered, then superconducting circuits consume considerably less energy. We discuss several superconductor-based neuromorphic hardware in the subsequent discussion.

*III.A.1. JJ-based Neurons and Synapses*

In a JJ, two superconducting regions are separated by an insulating region through which cooper pair tunnels from one superconducting region to the other [62]. Ballistic transport of digital signal can be passed through this tunneling junction with ultra-high speed (~ 1/100$^{th}$ of the speed of light) [63]. Above a sufficiently high current, ($I_C$), the junction becomes resistive, and voltage drop begins to appear across the junction. The switching from superconducting to resistive state occurs in ps timescale. The ultra-fast switching speed (~ 1.5 ps) of JJ and simple architecture makes it suitable for various types of design consideration. When the phase difference of the two superconducting regions whirls over once, a magnetic pulse is produced which is known as SFQ [64]. The flux has consistent shape and duration. Using these virtues of JJ, several neurons and synapses have been proposed.

Crotty *et al.* [39] proposed a current biased JJ-based spiking neuron in 2010 (fig. 4(a)). Here, two JJs combinedly work to produce neuronal spiking. These JJs are referred to as control junction and pulse junction. The neuron loop is connected to another loop that mimics electrical and chemical synapse [Fig. 4(a)]. When a bias current ($I_{bias}$) is applied, it is divided to both the JJs. $I_{bias}$ keeps both the JJs just below critical current. If an input current ($I_{in}$) with sufficient amplitude is applied, the current through the pulse junction exceeds its $I_c$, which injects a magnetic flux into the loop. This magnetic pulse creates a voltage drops across the pulse junction which creates the action potential (AP). As the flux builds, the current through the control junction reaches $I_C$ which eventually drains the flux in the loop. This way the pulse junction retains its initial superconducting state and gets prepared to fire again in the next cycle. The firing rate of this design is reported to be $2.0 \times 10^{10}$ AP/(neuron/s). It is theoretically shown that two individual neurons can be coupled together to implement both excitatory and inhibitory synaptic coupling depending on the sign of the bias current. For positive (negative) bias current, the second loop works as an excitatory (inhibitory) synapse. The inherent energy consumption of the circuit (without considering the cooling power) has been measured by the SFQ pulse of each of the JJs which is 70 zJ, 35 zJ, and 150 zJ for the input neuron, synapse MJJ, and output JJ.

**Table I:** Structural comparison of existing JJ-based neuromorphic hardware.

| Design | Input Neuron | Output Neuron | Synaptic Element | Synapse Type |
|---|---|---|---|---|
| Crotty *et al.* [39] | Current biased JJ loop | Current biased JJ loop | RC resonant circuit | Binary |
| Schneider *et al.* [64] | Current biased JJ | Current biased MJJ | Dynamically reconfigurable MJJ | Graded Weight |
| Schneider *et al.* [65] | Current biased JJ | Current biased SQUID | Dynamically reconfigurable MJJ | Graded Weight |
| Schneider *et al.* [66] | SQUID with resistive element | SQUID with resistive element | MJJ | Graded Weight |
| Goteti *et al.* [67] | Multiple JJ structure | Multiple JJ structure | SQUID loop | Graded Weight |

Although Crotty et al. [39] demonstrated excitatory and inhibitory synaptic behavior, their design does not contain the implementation of multistate synaptic weight and their dynamic tunability. Schneider et al. built a neuromorphic element where both non-magnetic and magnetic JJs were utilized and the critical current of the magnetic Josephson junction (MJJ) is dynamically tunable [64]. Here, only a current biased JJ is used as a neuron element. The MJJ is kept in the middle of two neurons, which weights the transmitted pulse from the input neuron to the output neuron (fig. 4(b)). The preneuron JJ produces spikes that are weighted in the MJJ. The critical current of this MJJ (from 1 µA to 100 µA) is analogous to the synaptic weight in a neuromorphic network and can be dynamically tuned by changing the superconducting order parameter [65]. The output neuron has a higher critical current (100 µA to 500 µA) and can fire based on the critical current of synaptic MJJ. This design reports a firing frequency of 0.35 GHz. The total energy for a single classification task is roughly 2 aJ (= 2 fJ with refrigeration) which is lower than the energy required for a single human synaptic event (10 fJ). The proposed JJ-based circuit exhibits neurosynaptic behavior with minuscule energy. However, $I_C$ modulation can be practically challenging due to the crosstalk between MJJ synapses.

In the subsequent work, Schneider et al. utilized a nanotextured MJJ as a synapse where a zero-field (0 T) reproducible analog control is demonstrated for the first time [65]. Here, they used a Si layer containing Mn nanoclusters as the tunneling barrier. The magnetic order of the Mn nanoclusters was tuned to vary the $I_C$ of the MJJ. The $I_C$ of the disordered state is higher for the disordered state than for the ordered state. To increase the synaptic weight (decrease the magnetic order of the nanocluster), electrical current pulses (4 pJ and 1 ns duration) at zero applied magnetic field need to be applied. However, to decrease the synaptic weight (increase magnetic order), current pulses were applied in the presence of an external magnetic field (~5 T). The number of electrical pulses and the pulse energy can finetune the critical current. Here, the energy for a JJ synapse with an elliptical cross-sectional area of $1.5 \times 3.5$ µm$^2$ is reported as 3 aJ [65]. Contrary to the previous work, here the post neuron is inductively coupled to the synaptic branch with a 50% coupling strength and this coupling works as a DC to SFQ converter (fig. 4(c)). Also, the post-neuron is designed by a current biased MJJ-based SQUID instead of a current biased JJ in the previous design. This inductively coupled structure reduces unwanted crosstalk between different circuit structures. Here, the estimated energy for a presynaptic spike weighted to generate a post-synaptic spike was reported as ~1 aJ. For a larger system (N ~1000), the energy would be ~1 fJ with the additional cost of cooling. Besides, SFQ-based circuits have been previously reported to work up to 60 gigabits/s from which it can be estimated to operate in hundreds of GHz range of frequency.

In their following work, Schneider et al. proposed their concept of coupling between multiple synapses with the neurons of the next stage in a network setting [60]. Here all the synapses are connected to a large output neuron by inductive coupling and the signals in the synapses are integrated into the neuron loop. The current through the JJ in the large neuron loop exceeds the $I_C$ based on the inputs of the synaptic circuits, and if it does, an SFQ pulse is created. The $I_C$'s of the synapse MJJs are modulated in a supervised fashion. This inductively coupled connection is designed to minimize unwanted cross-talks between inductively coupled circuit structures. In another work, the relationship of the inductively coupled current with the output SQUID neuron is analyzed and it is found that the coupled current level increases up to a certain value with the increase of the pulse duration but saturates after that point [66]. Goteti et al. have proposed a neural network with a disordered JJ array synaptic network where a JJ-based collective disordered synaptic network works with the neuron element to implement feedback loops for an unsupervised learning algorithm [67]. Here, a disordered array of n-different loops can have a 3n memory state.

To implement parallelism in a network, sufficient fan-in and fan-out capacity are required in a technology platform. Schneider et al. have reported a theoretical demonstration of a fanout of 1-to-10000 and a fan-in of 100-to-1 [68]. A rough estimate of the power dissipated for a 1-to-128 flux-based fan-out circuit for a given critical current value is reported to be 44 aJ. Both current-based and flux-based fan-in is analyzed, and it is demonstrated that for larger network signal current reduces significantly for current-based fan-in circuits. Therefore, for larger networks, the flux-based coupling is more efficient. The minimum diameter of a typical JJ is 500 nm [68] whereas MJJs for neuromorphic application have a cross-sectional area of a few $\mu m^2$. This is several orders of magnitude larger than the conventional CMOS-based designs [69]. However, for cryogenic memory applications, MJJ with a cross-section as low as 50 nm $\times$ 100 nm has been reported. Hence, cryogenic neuromorphic hardware based on MJJ is expected to reach a similar size. Although, miniaturization of MJJ size is an encouraging factor for area reduction, for an analog MJJ-based synapse, reducing the size will reduce the number of nanoclusters and thus the number of available synaptic states. Based on these design aspects, it is evident that JJ-based neuromorphic circuits show promising performance in their neuro-synaptic behavior in terms of speed and power efficiency. All the JJ-based designs require constant current biases with precise values. This is practically challenging. Moreover, the modulation of $I_C$ of the MJJ synapse requires a high external magnetic field necessitating additional hardware resources and adding to reliability concerns. Substantial research efforts are still required for realizing a JJ-based efficient and scalable neuromorphic network.

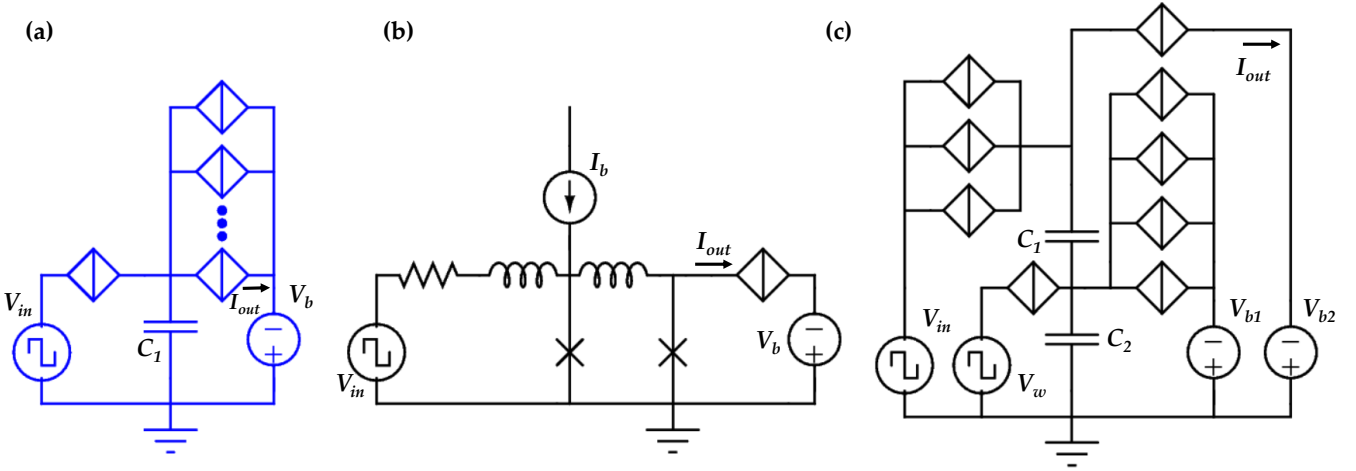

**Figure 5: (a)** QPSJ based neuron [74]. **(b)** QPSJ-based neuromorphic circuit having multi-synaptic state with both current and voltage bias proposed in [75]. **(c)** QPSJ-based neuromorphic circuit with multi-synaptic state with voltage bias only [76].

*III.A.2. QPSJ-based neuromorphic hardware*

Even though JJ-based designs reported so far exhibit neurosynaptic behavior with ultra-low power consumption, their designs include constant current biasing which lacks practical feasibility. QPSJ is an alternative superconducting device that comes into consideration for its voltage-based switching property. Quantum phase slip can be characterized as the dual process to the Josephson effect based on the charge-flux duality. In a QPSJ, flux-quantum tunnels across a superconducting nanowire along with the cooper pair transport and generates a voltage across the junction [70]. QPSJ has been reported to have a length as low as ~1 $\mu$m by Mooij et al. [71]. Besides, it is reported that a phase slip event can be expected in superconducting wires with a diameter as low as 10 nm [72]. The nm range of thickness indicates that QPSJ can be incorporated into a compact neuromorphic network. In a QPSJ, below a certain threshold voltage known as the critical voltage ($V_C$), no current flows due to the Coulomb blockade [73]. When an input voltage is above the $V_C$, the cooper pair begins to tunnel through the junction and the current begins to flow. Therefore, the I-V curve of a QPSJ is analogous to the V-I curve of a JJ. This switching characteristic of QPSJ generates spikes when a voltage pulse is applied across it.

By utilizing the switching behavior of QPSJ, Cheng et al. proposed a QPSJ-based spiking neuron for the first time in 2018 (fig. 5(a)) [74]. In their design, two branches of QPSJs are connected through a capacitor in between, which acts as an island of charge between them. Here, the first QPSJ is connected to the input. For an incoming input, the voltage across it rises. When the capacitance is smaller than a certain value ($C_1 < 2e/V_C$), it charges and discharges with the input spike. Thus, the neuron produces spike output for every spike input. This mode of operation is known as the tonic spiking mode. If the capacitance is above this value, the neuron works in tonic bursting mode. In this case, the input voltage pulses are integrated, and the output current spikes are fired once a certain threshold is reached. This design enables tunable firing frequency and threshold voltage. The second branch of QPSJs is kept at a bias voltage ($V_b$) to keep the branch of QPSJs just below $V_C$. The firing frequency can be varied by changing the series resistance of the QPSJ devices used. Besides, threshold voltage and firing rate both can be tuned by connecting several QPSJs in parallel (fig. 5(a)). To keep all the QPSJs equally biased, a resistor can be inserted, and the behavior perfectly matches with a LIF neuron behavior. The total firing energy for this circuit design is found by the switching energy of QPSJ multiplied by the number of QPSJ required for firing. In the parallel (second) branch, for only one QPSJ, the energy is 6.4 zJ (3.2 zJ × 2). This work reported only the spiking behavior of the neuron and the modulation of the spikes are achieved only by choosing the value of the capacitor between the two branches, hence, lacks dynamic tunability.

QPSJ has also been reported in novel neurosynaptic circuits in combination with JJ (fig. 5(b)). In the subsequent work, Cheng et al. combined the previously proposed QPSJ neuron with a synaptic circuit based on MJJ [75]. The tunable $I_C$ of the MJJ works as the synaptic weight for the device. This design does not consist of the island capacitor. When the current across the first JJ reaches the $I_C$, the current is bypassed to the second JJ and a voltage appears across it. This eventually increases the voltage across the QPSJ and switches it back to the superconducting state. Both a binary and multistate synaptic circuit has been proposed. In the binary synapse, an input voltage pulse can generate an output current spike based on the $I_C$ of the MJJ. Multi-state synaptic behavior is achieved by using two MJJs in parallel instead of one, where four different levels of the $I_C$ of the biased MJJ determine four synaptic states. Here, the synaptic weight ($I_C$) will determine the number of output spikes in a given period (2, 1, 0.5, 0). By increasing the number of JJs, the number of synaptic states can be increased.

A dynamically tunable neuromorphic network solely based on QPSJ has been firstly demonstrated by Cheng et al. (fig. 5(c)) [76]. Here, a QPSJ-based non-destructive readout memory circuit is used as a synapse. The charge capacitor island of fig.5(a) is replaced by two series capacitors ($C_1$ and $C_2$). The bottom capacitor ($C_2$) controls the number of cooper pairs tunneling to the output QPSJ. Therefore, the charge stored in $C_2$ acts as a synaptic weight for the circuit. Both binary and multistate synaptic memory state has been demonstrated by simple augmentation of the structure (fig. 5(c)). These cooper pairs give rise to the output current ($I_{out}$) of the design.

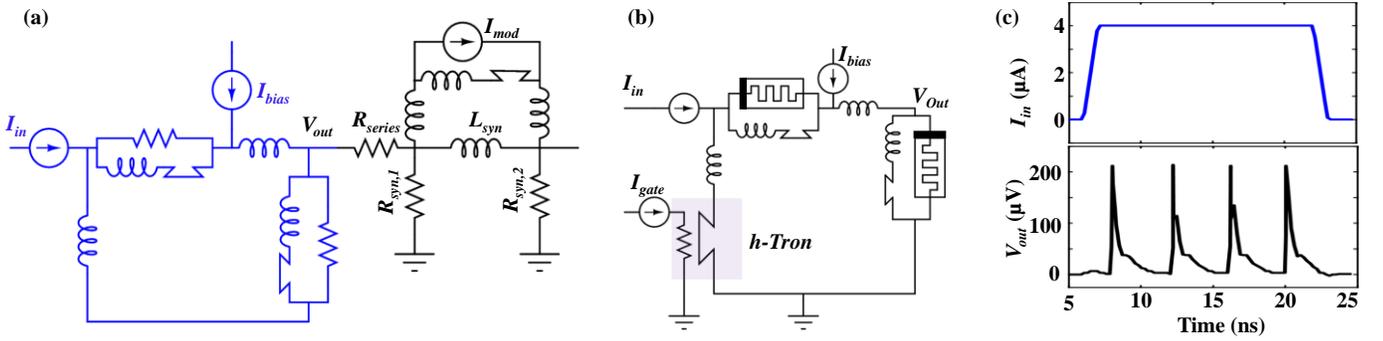

**Figure 6: (a)** Superconducting nanowire-based neuron (*blue*) and synapse (*black*) [81]. **(b)** Superconducting memristor-based neuron topology [82]. **(c)** Input Current ($I_{in}$) and corresponding Voltage spike ($V_{out}$). **(c)** Input Current ($I_{in}$) and corresponding Voltage spike ($V_{out}$).

The power required per switching event of a QPSJ-based neuron is estimated to be ~3.2 zJ which is orders of magnitude lower compared to a JJ-based neuron. Therefore, for a single firing event, a QPSJ-based neuron (with N=10 parallel QPSJ and a series QPSJ at the input) dissipates a power equivalent to ~35.2 zJ. The firing speed is estimated to be 10 spikes/ns which implies that the switching time is faster compared to JJ-based neurons. However, the development of a QPSJ exhibiting all the theoretically predicted behavior is still underway. Therefore, QPSJ is lagging behind JJ in terms of technological maturity. Although a QPSJ-based network has not been experimentally realized yet, it is expected to provide a lower chip area compared to the JJ-based neuromorphic hardware system as a superconducting nanowire has a diameter of several nm only [77]. Although the writing efficiency of the MJJ is high, the process requires an external magnetic field which is not suitable for real-time learning circuit operation. In this regard, electrically tunable QPSJ has the potential for more practical feasibility.

### III.A.3. Superconducting Nanowire-based Neuron and Synapse

For a large neuromorphic network, the number of SFQ pulses generated should be high enough to drive large fanout. In this case, JJ has limitations in terms of the number of SFQ pulse generated. So, it can be highly difficult to implement a full neuromorphic network solely based on JJ. Besides, the action potentials in JJ are not strong enough to be detected easily. One possible alternative of the JJ is thin superconducting wire, also known as superconducting nanowire (SNW). SNW exhibits reliable switching from superconducting to resistive state and they have been reported to generate a higher number of SFQ pulses as output [74]–[76], [78].

In an SNW, superconductivity breaks down and the wire turns resistive when the amount of current exceeds a threshold value referred to as the critical current ($I_C$). If the current is decreased below a certain level known as re-trapping current ($I_r$), the nanowire retains its superconducting state (fig. 3(c)) [79]. Typically $I_r < I_C$ and a hysteresis is observed due to this asymmetry. If the nanowire is placed in parallel with a shunt resistor, electrothermal interaction between the nanowire and the shunt resistance takes place [80]. Under a sufficient current bias, this interaction behaves like a relaxation oscillator. When the current exceeds the $I_C$, the nanowire turns resistive, and the current abruptly begins to flow through the shunt resistor. This increases the voltage across the shunt resistor, and we get a voltage at the output. Meanwhile, the current across the nanowire decreases as well. At one point the current across the nanowire goes below the $I_r$ and the nanowire becomes superconducting again. Therefore, the current through the shunt resistance drops and the voltage goes through relaxation. This way, an SNW shunted with a resistor generates spontaneous oscillation under a current bias with sufficient strength.

The unique relaxation oscillator discussed so far has been utilized by Toomey *et al.* to design a superconducting nanowire-based neuron circuit for the first time. The proposed neuron has the same circuit topology as the JJ-based neuron proposed by Crotty *et al.* (fig. 6(a)) [81]. The JJs are replaced by the SNW-based relaxation oscillator that we have just discussed. The interplay between the two oscillators generates spiking output voltage. Here, a current pulse in the input ($I_{in}$) produces a pulsating output voltage (fig. 6(c)). The bias current ($I_{bias}$) is chosen such that, it keeps the current flowing through both the nanowires just below the critical value ($I_c$). When $I_{in}$ is applied, it is added to the component of $I_{bias}$ in the nanowire of the main oscillator. Contrarily, the component of $I_{in}$ contradicts the existing $I_{bias}$ component in the control oscillator. This causes the main oscillator to switch to the resistive state whereas, the SNW in the control oscillator remains superconducting. In this way, a counterclockwise current is injected into the loop. This, in turn, helps to switch the nanowire of the control oscillator which reduces the counterclockwise current in the loop. and helps the main oscillator to spike again (fig. 6(c)). Thus, without the control oscillator, the main oscillator would only be able to fire once. Here, the firing frequency depends on the magnitude of $I_{bias}$. The threshold voltage for firing can be set by giving input pulses at the input. A negative pulse sets the threshold voltage low, and a positive pulse does otherwise.

For the connection between two neurons, an inductive synapse is introduced between the neurons to control the downstream signal. The inductor gets energized by the rapid voltage spikes from the neuron. The effect of the inductive synapse can be excitatory or inhibitory depending on the sign of $I_{bias}$ applied to the upstream neuron. The synaptic weight/strength can be varied by changing the effective inductance of the synapse. For applying such variability in the inductive synapse, the superconducting nanowire is connected in parallel with the inductor. Since a nanowire's kinetic inductance increases with the increase of its bias current, an ideal current source is placed in parallel with the synapse inductor ($L_{syn}$). The overall parallel inductance of the synapse can be modulated by the modulating current ($I_{mod}$). Although such an implementation can be suitable for deep learning, it does not capture the true behavior of bio-realistic synapses. Here, the pulse energy is on the order of ~10 aJ. The neuron circuit consumes about 0.05 fJ energy for each action potential or firing event (without cooling energy), while for the synapse, the energy is around 0.005 fJ. Here, a superconducting transmission line (axon) is also reported to transport the neuron signals. The neuron signal that is passed through the axon acquires a delay of 100-500 ps which is the full width of an action potential. This duration is significantly shorter than the mammalian brain (a few ms). Owing to superconducting interconnects, the resistive loss is not present here and the nanowire itself consumes minuscule energy for each action potential (~0.05 fJ excluding the cooling power). The combination of these two leads to an unprecedented power efficiency. Besides, the nanowire-based neuron has been experimentally demonstrated where NbN nanowire with a diameter of 60 nm is used [77]. The total area of each oscillator is around 10.87 µm × 8.72 µm and the large inductor occupies the major portion of the area (around 5.81 µm × 2.95 µm). The total area occupied by a single neuron can be estimated to be around 27.17 µm × 25.64 µm. Hence, there is a significant area penalty in this demonstration. However, a small 9 ×3 sized network for a simple image recognition has been reported for image recognition application as a functional verification approach. Here, the authors have used the LTspice model of SNW and reasonable accuracy is achieved [77].

The design proposed by Toomey *et al.* offers limited reconfigurability in terms of spike modulation. The value of $I_{bias}$ controls the spike frequency and amplitude. However, $I_{bias}$ can be reliably varied within a very small range (~2 µV). Islam *et al.* proposed an SNW-based neuron circuit with higher reconfigurability (fig. 6(b)) [82]. Here, a superconducting memristor with two different resistance levels was used instead of a fixed resistor. The superconducting memristors is primarily a SQUID device which have two different JJs that has two different levels of conductivities [83], [84]. This asymmetry gives rise to the two different resistance levels (~60 mΩ and ~34 mΩ). Two different resistance levels offer two different oscillation frequencies. Since there are two oscillators in this topology, four distinct ranges of spike frequencies are achieved (from ~700 kHz to ~2.7 MHZ). The proposed neuron consumes (estimated) ~10 atto-Joule per spike (without considering the cooling energy). The primary drawback of this design is the level of spike voltage. As the resistance level is in the range of several tens of mΩs, the spike amplitude is in the range of several µVs which is experimentally challenging to detect. Furthermore, to program the superconducting memristors in different resistive states, a heater cryotron device (*h*Tron) is employed as an aid which requires an additional gate current ($I_{gate}$) to control.

SNW shows superior performance in terms of fanout in a neuromorphic network [77]. JJ and SNW both have quantized flux outputs. But SNW has a considerable advantage over JJ in terms of ease of fabrication, CMOS compatibility, and scalability. SNW permits higher degrees of parallelism compared to JJ and consequently, SNW permits a higher fanout. In addition to the fan-out and fan in, the nanowire output spikes have a high magnitude and duration making them compatible with the CMOS control circuitry. This is a clear experimental advantage over other superconducting devices. The quantum phase slip phenomenon in a nanowire is much more experimentally challenging than the resistive switching process. In this regard, the SNW has much higher reliability than the QPSJ. However, the SNW-based neuromorphic circuits require accurate constant current bias, similar to the JJ-based neuromorphic circuits, posing similar experimental challenges. Moreover, its applicability in a full-scale network is yet to be seen. Despite the scarce research efforts on SNW-based neuromorphic circuits, their promising characteristics incite further research exploration.

### III.B. Hybrid Superconductor-Semiconductor-based Neuromorphic Hardware.

Since the optical signal propagates at the speed of light, it is intuitive to give the effort to build a fast-neuromorphic network based on the optical signal [85]. Furthermore, optical signals do not contain capacitive and inductive attenuation, which makes it suitable for long-distance transmission. SNW can be optically sensitive and is used in superconductor-based photon detectors (SPD). Shainline *et al.* first proposed a hybrid semiconductor-superconductor hardware platform where a light-emitting diode (LED) with SPD work in combination to behave as a spiking neuron [85]. Here, electromechanically coupled optical waveguides connect the neurons and provide variable strength of connectivity. The distance between two waveguides can be changed by modulating voltage. A high tunability can be achieved through this electromechanical process and the coupling strength acts as a synaptic weight. For the detection of a photon, a superconducting nanowire single-photon detector (SNSPD) is used. The photon detection efficiency is > 90% keeping the attenuation negligible. For detecting and processing optical signals, SNSPD is a natural choice, thanks to its capability of waveguide integration, high speed compact size, and ease of fabrication [86], [87].

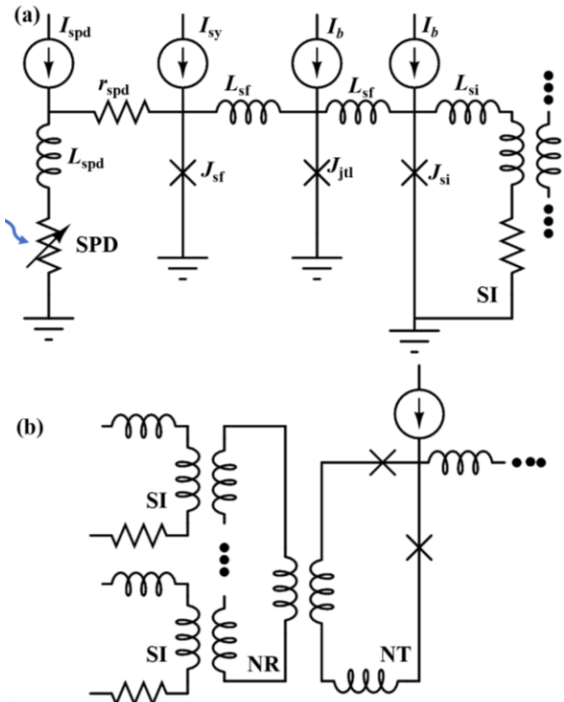

**Figure 7: (a)** Circuit diagram of a photon to fluxon transducer. **(b)** Circuit diagram of multiple synapses (SI loops) connected to the NR loop and the neuronal thresholding (NT) loop through mutual inductors [91].

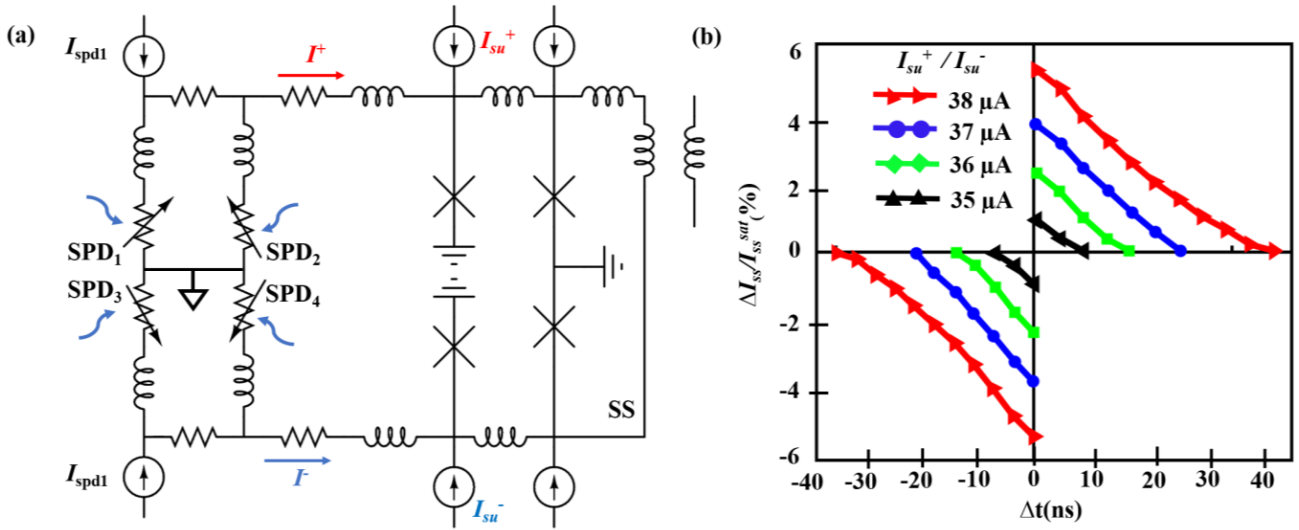

**Figure 8: (a)** Implementation of spike-timing-dependent plasticity. **(a)** Circuit under consideration [91]. **(b)** Change in current stored in the SS loop as a percentage of the total storage capacity of the loop vs the relative arrival time of the two photons involved, Δt.

An SNSPD has two primary components: nanowire detector and LED. SNSPDs can successfully sense individual photons and converts the optical signal to electrical output [88]. When SNSPDs are connected in parallel with an LED, the detector acts as a photonic signal integrator, and this type of detector is referred to as the parallel nanowire detector (PND). The LED has a finite nonzero resistance. In the absence of any incoming photon, the SNSPD remains superconducting. For a PND, if any of the SNSPDs remain superconducting, then all the bias current passes through it. However, when all the detectors in a PND detect photons, they all turn resistive and a non-zero net resistance emerges. In another word, the bias current exceeds the critical current of the PND and is redirected through the LED. This current generates a light signal which propagates to the subsequent neurons through the waveguides. It is also possible to connect two SNSPDs in series so that absorption in one of the SNWs, decreases the firing threshold of another one. This kind of arrangement is known as the series nanowire detector (SND). In such a network, absorption in the upstream nanowire decreases the current flowing through the branch and thus more photons are required to be absorbed by another nanowire connected in series [89].

The optical neurons discussed so far contain LEDs that are connected in a simple arrangement without any optical gain. The bias current value has to be compatible with both the SNSPD and the LED. It is possible to decouple the firing threshold and LED gain by introducing a cryotron device. There are two types of cryotron devices, nano cryotron (nTron) and heater cryotron (hTron). The nTron is a three-terminal device that works as a switch owing to the joule heating process. When the gate current of an nTron exceeds a critical current, the path from source to drain becomes resistive. If connected with a load parallelly, the current is bypassed through it. Shainline *et al.* utilized this characteristic to drive the output LED to be driven by a certain bias current [85]. In this work, individual neurons are connected through optical waveguides. The optical waveguide-based connection has a distinct advantage over the electrical connection. It allows individual neurons to collect and integrate optical signals from many neurons and integrate them without time-multiplexing [90]. For instance, a stingray neuron has been proposed where each neuron combines the modes from many waveguides without significant loss via a landing pad containing the PND arrays. Hundreds of waveguides are combined in a receiver body with less than 0.2 dB insertion loss from any port.

As a synaptic weight, here electromechanically actuated waveguide couplers are proposed for the network. Here, as well, the coupling can be varied by varying the distance between the waveguides electromechanically by applying voltage. The maximum distance (minimum coupling) occurs at 0 V between the two waveguides. Upon activity, a voltage difference is created between the two waveguides which increases the coupling strength. Spike Timing Dependent Plasticity (STDP) can be implemented if closely spaced firing events from upstream and downstream neurons place charges on the capacitor.

There are three primary sources of energy dissipation in this design, the inductive energy of supercurrent flowing through the circuit, the capacitive energy of the LEDs, and the energy required to produce the photons. Here the energy required for each photon generation depends on the LED efficiency. Most of the energy is consumed in supplying current to the inductors and charging the capacitors. At unity efficiency, the energy required is 2 aJ/photon while for 1% efficiency the energy becomes 20 aJ/photon which is orders of magnitude lower compared to the CMOS-based neuron. Thus, the use of a hybrid platform of superconducting electronics and optoelectronics offers a factor of $10^6$ improvements in energy efficiency compared to the best-performing CMOS process available.

In the subsequent work, Shainline *et al.* proposed a slightly different topology of optical neuron referred to as superconducting optoelectronic loop neuron [91] (fig.7). Here, as well, light is used to communicate between neurons. However, the synaptic weight is implemented in the electric domain. A neuron with a single integration loop is capable of receiving inputs from thousands of synaptic connections. In this work, a superconducting photonic detector is used in conjunction with a JJ to detect photons and then integrate this current through a synaptic integration (SI) loop (fig. 7(a)) [92]. With a single photon detection, the detector transduces the photon to an electrical current. The current through the JJ exceeds the $I_C$ and produces flux pulses known as fluxons. Here, fluxons are voltage spikes across the synaptic JJ and they are integrated as the current in the synaptic integration loop. The neuronal receiving (NR) loop integrates the current that it receives from all the neuron's synapses (fig. 7(b)).

In this design, the use of mutual inductance allows a higher number of synapses to relate to one neuron. Besides, the NR loop is devoid of any current leakage path (fig. 7(b)). The NR loop is connected to a larger loop referred to as the neuronal thresholding (NT) loop (fig. 7(b)). This NT loop works as a transformer that steps up the current to be detected beyond the threshold value. Here the fluxons are referred to as synaptic firing events and the number of synaptic firing events can be controlled by a current biased JJ ($J_{sf}$). The critical current of $J_{sf}$ can be varied by a current bias and it can control the number of fluxons that are created. For low critical current, the SPD current is higher than the critical current for a long duration and thus it creates many fluxons. If the critical current is high, the number of fluxons is lower. This critical current is the synaptic weight for this circuit. This topology also offers excitatory and inhibitory connections between upstream and downstream neurons. The sign of the mutual coupling between the Synaptic Integration (SI) loop and Neuronal Receiving (NR) loop determines the excitatory and inhibitory mode of operation in this topology [91].

To vary the synaptic strength, the synaptic current ($I_{sy}$) can be dynamically tuned (fig. 7(a)). Shainline *et al.* reported variable synaptic strength applicable for both supervised and unsupervised learning [91]. To implement supervised learning, two additional control pulses are used ($I^+/I^-$) which can turn the synapses either into the potentiation or depression state of operation (fig. 8(a)). To modulate the synaptic weight in smaller steps, a superconducting loop is proposed that can store multiple fluxons simultaneously. Here a DC-to-SFQ converter is used to add flux quanta one by one. Two separate biasing signals $I^+$ and $I^-$ can drive the synapse to either potentiation or depression mode by adding fluxons in opposite directions [91].

As a further extension of the work, the Hebbian learning rule is also realized by designing two correlated detection schemes. When a correlated SNSPD detects a photon from post-neuron within a certain time after the pre-neuron fires, a current is injected into the synaptic storage loop. The amount of current that is injected into the synaptic storage loop depends on the time difference between the firing events of the pre-neuron and the post-neuron. Spike-timing dependent plasticity (STDP)-based learning is also reported by mirroring the Hebbian learning circuit for a current injection in the SS loop from the opposite direction. Here, synaptic depression occurs when the presynaptic neuron fires later than the post-synaptic neuron (fig. 8(a-b)) [93]. When a photon is detected in the pre-neuron detector closely followed by a photon detection in the post neuron, $I_{spd1}$ is diverted to the synaptic storage loop to increase the injected current in the clockwise direction. This increases the synaptic weight. In the mirror circuit, SPD3 and SPD4 detect photons from post-neuron and pre-neuron respectively (fig. 8(a)). Consequently, when the post neuron fires a photon before the pre-neuron, a current is injected in the SS loop in the counterclockwise direction, and this pushes the current in the counterclockwise direction decreasing the synaptic weight. Thus, the synaptic weight can be modified by the timing difference of pre-neuron and post-neuron firing. The SI loop is coupled with the synaptic biasing loop that determines the biasing current of the synapse. The biasing current is used to inject current in the synaptic integration loop which is coupled to the neuronal thresholding loop. Current from the neuronal thresholding loop is sent to the cryotron-based transmitter circuit to produce an optical signal via LED.

A larger network of superconducting-optoelectronic hybrid platforms has been reported in [91]. Here, the estimated energy consumption and area footprint show promising prospects. The switching energy for an SNSPD is reported to be 21 zJ. The estimated power consumption is 1 mW for a network with 8100

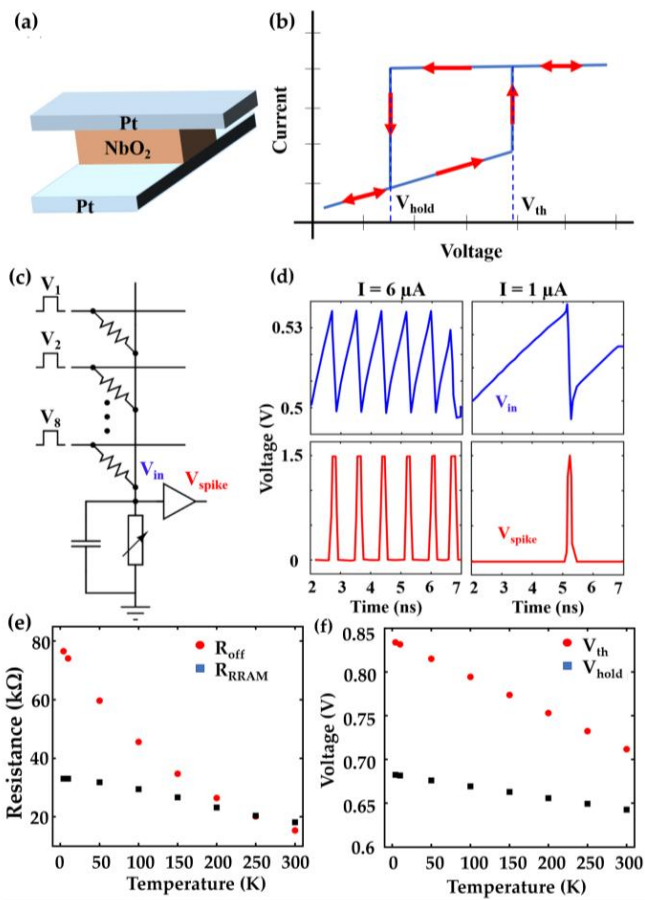

**Figure 9 (a)** A Pt-NbO$_2$-Pt device with its **(b)** *I-V* characteristics. **(c)** Proposed design of an oscillation neuron with a MIT device at the end of the column that emulates the $V_{in}$ node [56]. **(d)** The waveform of the CMOS integrate-and-fire neuron for different weighted sum current values (6 μA & 1 μA). **(e)** variation of RRAM resistance and off-resistance with temperature. **(f)** The threshold voltage ($V_{th}$) and the hold voltage ($V_{hold}$) variation with temperature.

neurons and the total on-chip power is less than 10 mW. This includes the power consumed by the biasing currents [91]. The reported area of a network with 8100 neurons including three different pairs of waveguide routing is about 1 cm$^2$. The overall area of such a network depends on the degree of interconnectivity nodes. Although the inductor in the SI and SS loop is reported to have a value of 1 $\mu H$ with an area of 5 ×5 $\mu m^2$, the most area-consuming component of the neuron is the driver amplifier for LED and mutual inductance. But the overall area is mostly dictated by the passive waveguide requirement. In this platform, a system with $10^6$ neurons and $200 \times 10^6$ synapses on a 300 mm wafer consume 1 W of power including the cooling power. This energy specification indicates that this platform is very promising in large-scale network implementation. The network offers tunability of synaptic weight, neuronal threshold, and firing frequency and can implement the STDP-based learning rule via the current biasing scheme. Table II summarizes the relative comparison of the fan-in and fan-out capacity of different superconducting neuromorphic devices. Because optical interconnect has no resistance, capacitance, or inductance, they can provide massive fanout capacity as shown in Table II. The hybrid semiconductor-superconductor platform has a better fan-in and fan-out capacity and consequently offers a higher parallelism advantage compared to the solely superconductor-based neuromorphic hardware platform.

### III.C. Non-Superconducting Cryogenic Materials

Several materials that exhibit threshold switching behavior, have been shown to preserve their characteristics in cryogenic temperature. Hence, they were explored for efficient neurosynaptic behavior in cryogenic temperature. Certain underlying physics can be attributed to the switching behavior in these materials. For example, nonlinear transport mechanism such as the Mott transition has engendered unique advantages in an NbO$_x$-based material [94], [95]. A simple stack of a Pt-NbO$_2$-Pt is shown in fig. 9(a). When a sufficient voltage ($V_{C\text{-}IMT}$) is applied in the NbO$_2$-based device, it undergoes the insulator to metal transition. During the reduction of the voltage across it, the material retains its insulating state below a certain voltage ($V_{C\text{-}MIT} < V_{C\text{-}IMT}$). This switching behavior manifests hysteresis in its I-V characteristics which have been utilized in designing novel neuromorphic circuits (fig. 9(b)).

### III.C.1. Threshold switching-based neuron

Several oxides (HfO$_2$, VO$_2$, NbO$_2$) have been reported to exhibit metal to insulator transition and abrupt switching characteristics in low temperatures. Fang *et al.* have analyzed the cryogenic behavior of HfO$_x$-based synapse and demonstrated that HfO$_x$ can maintain its switching behavior at 4K [96]. Both $R_{HRS}$ and $R_{LRS}$ decrease with increasing temperature in the case of a Pt/HfO$_x$/TiN structure and high $R_{HRS}/R_{LRS}$ is observed for low temperature which is a potential requirement for RRAM-based synapse.

When NbO$_x$ is connected with an external resistor, the voltage across the NbO$_x$ (membrane voltage) starts to self-oscillate, and here the oscillation frequency is proportional to the conductance (fig. 9(d)). For an RRAM connected in series with it, the oscillation frequency is proportional to the RRAM conductance, and therefore it is feasible for integrating the weighted sum current in a crossbar synaptic array (fig. 9(c)). From the graph, it is evident that NbO$_2$ manifests hysteretic behavior within a large range of temperatures and hence, is suitable for the cryogenic environment (fig. 9(f)). Wang *et al.* have analyzed a NbO$_2$-based oscillatory neuron in series with previously proposed HfO$_x$-based cryogenic RRAM-based synapse [55]. They demonstrated that NbO$_2$ maintains its hysteretic behavior at a very low temperature (fig. 9(e)). Beyond the threshold voltage ($V_{th}$), it switches to a low resistive state (LRS), and below a certain voltage ($V_{hold}$), it retains its high resistive state (HRS) (Fig. 9(b)). When a voltage is applied, the parasitic capacitance C$_1$ gets charged by the voltage across NbO$_2$. If the voltage is greater than the threshold voltage, then it switches to the LRS. Now if the voltage across the neuron is less than $V_{hold}$, then the neuron goes back to the HRS, and C$_1$ is discharged. This way the neuron oscillates between the two voltage levels, $V_{th}$ and $V_{hold}$. From the temperature frequency analysis, it is seen that, even though the oscillation frequency increases with temperature, the oscillation frequency at an ultra-low temperature (4 K) is still in the GHz range.

To control the oscillatory behavior of the NbO$_X$-based device, Chen *et al.* have derived a performance control method for the oscillation behavior of the NbO$_2$-based devices very recently [30]. In this Pt-NbO$_x$-TiN-based device, a threshold switch is fabricated. The NbO$_x$ layer is divided into two parts, a NbO$_2$-based switching layer and a conducting layer of Nb$_2$O$_5$. Both $V_{th}$ and $V_{hold}$ increase at low temperatures. But $V_{th}$ increases more than $V_{hold}$ (fig. 9(f)). So, an expansion of the hysteresis loop occurs with the decrease of temperature. Switching time is another way by which we can modulate the oscillation frequency. As the temperature decreases, both the switching time and thermal conductivity increase. Thus, the oscillation frequency decreases with the temperature. It is evident that at low-temperature NbO$_2$ exhibits high $R_{OFF}$ and exhibits superior performance. The higher the threshold voltage, the oscillation frequency is lower for a particular switching time. The fabricated Pt-NbO$_x$-Pt device has an active area of 10 ×10 μm$^2$ in a cross-point architecture. This is ~13 times smaller than the equivalent CMOS-based architecture [97]. It is evident that at low-temperatures, NbO$_2$ exhibits high $R_{OFF}$ and therefore superior performance. Fig. 10(b) summarizes the characteristics of the different cryogenic platforms and their broader advantages.

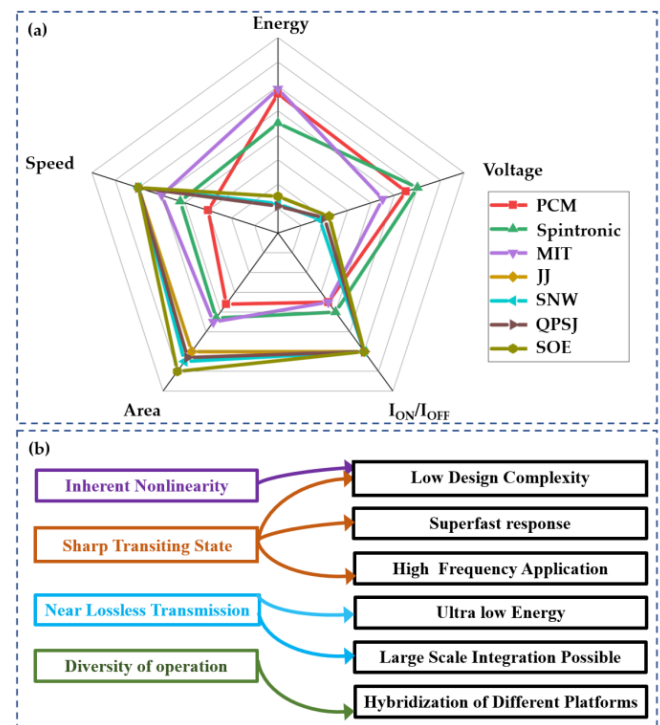

**Figure 10:** **(a)** Comparison of several existing platforms and cryogenic hardware platforms on five different metrics. **(b)** Features of the cryogenic devices and their corresponding advantages for neuromorphic hardware.

Table II: Comparison of different cryogenic neuromorphic hardware platform.

| Device Type | Structure | Operating Voltage | Temperature | Number of States | Firing Energy | Spikes per second | Ref. |
|---|---|---|---|---|---|---|---|
| Josephson Junction Based | JJ (Nb/Mn/Nb/SiO$_2$) as synapse | 150μV | 4K | Number of $I_C$ of MJJ | 150 zJ (SFQ pulse energy) | 1.6×10$^9$ spikes per second(est.) | Schneider *et.al.*, 2017 [66] |
| Quantum Phase Slip Junction (QPSJ) based | Mo–Ge deposited on suspended carbon nanotubes | 1mV | <1.92K | Both binary and multiple states are possible. Modulated by critical voltage of QPSJ (Cheng *et. al.*, 2021) | 6.4 zJ (switching energy × N) | 2×10$^{10}$ spikes per second (estimated) | Cheng *et.al.*, 2018 [74] |
| Superconducting Nanowire Based | NbN/SiOx | 250μV(est.) | Cryogenic Temperature | Depends on the number of the modulating current bias | 50 aJ (switching energy) | 0.25×10$^9$ spikes per second(est.) | Toomey *et.al.*,2019 [77] Toomey *et.al.*,2020 [78] |
| SNSPD based Neuron | NbN nanowire as SNSPD, NbN on GaAs substrate as waveguide. | Not reported | 2K | Multiple | 20 aJ / photon 20 fJ / photon (including cooling) | 10$^{15}$ synaptic event per second | Shainline *et.al.*,2017 [85] |
| Superconducting Optoelectronic Loop Neuron | NbN nanowire as SPD | >0.1V for LED to generate photon (Shainline et.al.,2017) | 4.2K | Multiple (a few hundred) | 2.7 aJ (weak synaptic efficacy) 41aJ (strong synaptic efficacy) | 10$^{15}$ synaptic event per second | Shainline *et.al.* 2019 [91] |
| Pt/NbOx/TiN device | Pt/NbOx/TiN on TiN/ SiO2/ Si substrate | 2V | 100K-300K reproted | Binary (RRAM) | Not reported | Spiking rate Depends on Temperature (~14×10$^6$ to ~7×10$^6$) | Chen *et.al.*,2021[30] |

## IV. Comparative Study

In previous sections, we discussed state-of-the-art cryogenic memory devices. In this section, we compare them based on energy consumption, spike generation rate, and several other metrics. The existing neuromorphic platforms such as CMOS-based, memristor-based, PCM-based, Spin-based, and FeFET-based are technologically more mature than cryogenic neuromorphic hardware. Some CMOS-based digital implementations of neuromorphic chips have already been fabricated by Intel and IBM[98], [99]. However, the design requirement is still strenuous, and the power consumption is still far beyond the power consumed by the biological counterpart. Although there exists a power disadvantage of refrigeration in cryogenic neuromorphic hardware, the intrinsic power consumption is still far below the conventional platforms. The resistive interconnect leads to the supply voltage degradation and self-heating in a conventional crossbar-based synaptic network. On the contrary, dissipation-less superconducting interconnect in cryogenic temperature provides a pathway to curtail the interconnect loss. Still, much more research is required for superconducting interconnects to match the massive parallelism of the human brain. Besides, most cryogenic devices are based on superconducting entities (JJ, QPSJ, SNW). JJ-based neurons and synapses operate in the μV range. Due to the low operating voltage, JJ-based neurons consume several attojoules (aJ) of power intrinsically. Even with the additional energy required for cooling, the overall power consumption remains orders of magnitude lower than the conventional platforms [100]. Another important aspect of the neuromorphic system is the design of analog-to-digital conversion (ADC) blocks. This is out of the scope of our review. ADCs are known to be the most power-hungry and area-consuming modules in conventional crossbar-based neuromorphic systems. The possibilities of (i) designing specialized area-efficient cryogenic ADCs or (ii) averting the need for ADCs by implementing a fully analog neuromorphic system are yet to be thoroughly tested/validated. There have been several reports on superconducting ADCs with promising figures of merit [101]–[103]. However, this is still an evolving field of research and there is a strong need for further research exploration and scrutiny.

For QPSJ-based neurons and synapses, the power consumption is lower (several zJ) than that of the JJ-based neurons and synapses. QPSJ-based multi-state synapses have also been theoretically proposed which consume zJ range of power. Moreover, the spiking rate is on the order of $10^{10}$ s$^{-1}$ for QPSJ- based neurons. For nanowire-based neurons, the power consumption is reported to be 0.05 fJ. Besides, it is presumed that the nanowire-based neurons are capable of a high fanout network due to the production of a massive amount of flux quanta. However, the exact fanout capacity is not quantified. Although optoelectronic-based neurons and synapses require semiconductor and superconductor hybrid implementation, they are capable of faster communication between neurons and synapses because they transmit signals at the speed of light. Due to the high number of synaptic events, they are suitable for a network with a higher degree of parallelism.

Finally, $NbO_2$-based neuron is promising for their natural oscillatory behavior, temperature-dependent spiking rates, and wide range of operating temperatures. Efforts have also been given to utilize the advantages of several devices by combining them into a single circuit. Table II summarizes the overall performance comparison of the state-of-the-art cryogenic neuromorphic hardware.

## V. Conclusion

With the ongoing pursuit of artificial intelligence hardware, neuromorphic architecture is the most promising among them, thanks to its inherent in-memory computational dynamics and massive parallelism. But the power efficiency and speed of the existing neuromorphic hardware platform are yet to be comparable with the human brain. Interestingly, cryogenic devices can circumvent the challenges faced by their conventional and semiconductor counterparts. These devices have been the center of attention in the scientific community for many years and recently, they have been employed to design neuromorphic hardware with an unprecedented performance metric. However, a comprehensive overview is still needed to summarize the state-of-the-art cryogenic neuromorphic hardware technologies and to meet the curiosity of future research enthusiasts. In this review, we have carefully classified the existing cryogenic neuromorphic devices and thoroughly discussed their advantages and disadvantages. Although there are cryogenic versions of conventional CMOS devices, they suffer from performance degradation in ultralow temperatures [104]. We have not covered that in our review as cryogenic CMOS itself is another wide area of research. Possibly, superconductors and superconducting devices are the most viable technology platform in cryogenic temperature. The recent demonstration of a superconducting processor has been proved to be ~80X less than that of its semiconductor counterpart (considering the cooling cost) [105]. However, superconducting devices suffer from scalability issues due to the coherence length (~200-700nm) constraints and large size. On the other hand, CMOS technology is highly mature and scalable. Nevertheless, superconducting electronics offer excellent speed and power consumption performance compared to any other of their conventional counterparts [84], [106], [107]. Figure 10(a) shows a relative comparison of some cryogenic devices with a few leading conventional platforms for implementing neuromorphic hardware. Figure 10(b) shows several properties of cryogenic device platforms and their corresponding advantages in a neuromorphic network. From our review, it is evident that cryogenic platforms are superior to conventional platforms in several important performance metrics. Undoubtedly, it is in our best interest to prepare ourselves for further exploration of cryogenic neuromorphic hardware to enhance its performance and achieve a more complete technology platform.